\documentclass[aps,pre,twocolumn,superscriptaddress,nofootinbib,longbibliography]{revtex4-2}

\usepackage[T1]{fontenc}
\usepackage[utf8]{inputenc}
\usepackage[english]{babel}
\usepackage{xcolor}

\usepackage{amsmath,amssymb,amsfonts,bm,bbm,mathtools}
\usepackage{physics}
\usepackage{microtype}
\usepackage[colorlinks=true,linkcolor=blue,citecolor=blue,urlcolor=blue]{hyperref}

\usepackage{tikz}
\usetikzlibrary{arrows.meta,calc,positioning}

\newcommand{\C}{\mathcal{C}}
\newcommand{\Ch}{\mathcal{C}} % added for compatibility with \Ch1 used in the text
\newcommand{\ii}{\mathrm{i}}
\providecommand{\dd}{\mathrm{d}}
\newcommand{\ee}{\mathrm{e}}

\begin{document}

\title{Discrete Action, Graph Evolution, and the Hierarchy of Symmetries:\\
A Rigorous Construction of Temporal Layers $\C1\!\to\!\C2\!\to\!\C3\!\to\!\C4$}

\author{Medeu Abishev}
\affiliation{Al-Farabi Kazakh National University, Almaty, Kazakhstan}
\affiliation{National University of Singapore, Singapore}
\author{Daulet Berkimbayev}
\affiliation{Al-Farabi Kazakh National University, Almaty, Kazakhstan}
\date{\today}

\begin{abstract}
Postulating a minimal discrete quantum of action $S=\hbar$ and a simple rule for the growth of an oriented graph, we construct a strict hierarchy of temporal layers $\C N$ with discrete periods $\tau_N=N\hbar/E$. Each layer is specified by its configuration space, symplectic structure, update rule, and emergent symmetry. At $\C1$ the state is represented by a single oriented edge with $U(1)$ phase $\ee^{\ii E t/\hbar}$. The transition $\C1\!\to\!\C2$ splits the edge into two independent flows, which yields canonical pairs $(x_a,p_a)$, local $U(1)$ invariance, and an effective $(2{+}1)$ metric with signature $(+--)$. The closure $\C2\!\to\!\C3$ produces $SU(3)$ connections and an Einstein–Yang–Mills type action. We show that these structures follow from discrete-action principles, and that stochastic graph growth naturally provides mechanisms for decoherence and spontaneous symmetry breaking.
\end{abstract}

\maketitle


%========================================
\section{Introduction}
%========================================

\noindent In modern physics, evolution is usually treated as a continuous process on a differentiable manifold: from Newtonian mechanics to the Schr\"odinger and Einstein equations \cite{Newton,Schrodinger1926,Einstein1915}. Despite the broad success of this viewpoint, there are indications of indivisible quanta of dynamics or granular bounds at microscopic scales.

A natural reference scale is the Planck time $t_{\mathrm{Pl}}\!\sim\!5.4\times 10^{-44}\,\mathrm{s}$ \cite{Planck1900}. While it does not, by itself, assert the discreteness of time, the quantum of action $\hbar$ already constrains evolution via time–energy uncertainty and quantum speed limits \cite{MandelstamTamm,QSL_Review,InformationDynamics}. This motivates an operational stance: physically accessible updates occur at discrete ``ticks'', with the continuous time parameter serving primarily as an interpolation label between them.

Discrete microstructures have been explored in many programs—from causal sets and loop quantum gravity to triangulations and noncommutative geometry \cite{Bombelli1987,Dowker2005,CausalSets,AshtekarLewandowski,RovelliSmolinArea,CDT_Review,Regge1961,Connes1994,GFTReview}. Discrete-time evolution also appears in lattice gauge theory, quantum walks, and quantum cellular automata \cite{LatticeWilson,QWReview,QCA,tHooftCA}. Preserving relativistic symmetries imposes stringent constraints on admissible discretizations \cite{CausalSets,PolymerQuant,DSR_AmelinoCamelia2002,Mattingly,HossenfelderPhenomenology,AmelinoCameliaLIVObs}.

Here we propose a minimal, discrete operational framework in which $\hbar$ governs both dynamics and symmetry formation. Only state updates that accumulate a quantum of action are allowed. We show how effective continuum descriptions emerge when the tick is small compared to system timescales, with generators remaining regular without \emph{ad hoc} cutoffs \cite{DiscreteNumerics}. We outline phenomenological windows—from interferometric phase accumulation and quantum control to cosmology \cite{HossenfelderPhenomenology,AmelinoCameliaLIVObs}. The approach is related to causal sets, loop gravity, and ``quantum graphity'' \cite{LQG,PathIntegralGraphs}, but differs crucially: the discrete variable is the action increment attached to each new graph edge.

%========================================
\section{Discrete action and the graph rule}
%========================================

\noindent We work with graphs as a mathematical representation of the system rather than as physical objects. At the beginning of continuous time, consider a system specified by a single loop/edge with elementary discrete action
\begin{equation}
S_1 \equiv E_1 \tau_1 = \hbar, 
\qquad \tau_1 = \frac{\hbar}{E_1},
\label{eq:S1}
\end{equation}
where $E_1>0$, while $t\in\mathbb{R}$ remains a macroscopic label.

The hierarchy of layers is introduced inductively:
\begin{equation}
\tau_N = N\tau_1, \qquad S_N = N\hbar.
\end{equation}

Different directions for transporting action $S$ along the initial loop are allowed, encoded by a minimal ``spin'' $s$ (not identified with the quantum-mechanical spin). The oriented ``spin–orbital'' separation between $\{\ket{b},\ket{e}\}$ is captured by
\begin{equation}
    \Delta
    =\hbar\!\begin{pmatrix}0&1\\-1&0\end{pmatrix}
    =\ii\hbar\,\sigma_y,
\label{eq:Delta}
\end{equation}
so that $b\!\to\!e$ measures $+\hbar$ and $e\!\to\!b$ measures $-\hbar$. Spin conservation is equivalent to conservation of total angular momentum.

A system at level $\C N$ is an oriented graph $G_N$ with $N$ edges evolving under:
\begin{enumerate}
\item one new oriented edge per step;
\item conservation of total action: $S_{n+1}=S_n+\hbar$;
\item invariance of total $J_z$.
\end{enumerate}

%========================================
\section{Layer $\C1$: single edge and $U(1)$ phase}
%========================================

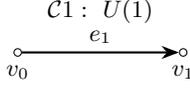
\begin{figure}[t]
\centering
\begin{tikzpicture}[scale=1.1,>=Stealth]
\node[circle,draw,inner sep=1pt,label=below:$v_0$] (A) at (0,0){};
\node[circle,draw,inner sep=1pt,label=below:$v_1$] (B) at (2,0){};
\draw[->,thick] (A) -- (B) node[midway,above] {$e_1$};
\node at (1,0.5) {$\C1:~U(1)$};
\end{tikzpicture}
\caption{Layer $\C1$: a single oriented edge between $(v_0,v_1)$ with $U(1)$ phase evolution ($e_1$).}
\label{LayerC1}
\end{figure}

\noindent Configuration: a single oriented edge $e_1:(v_0\!\to\!v_1)$, Fig.~\ref{LayerC1}. Phase space:
\begin{equation}
\mathcal{M}_1=S_1, \qquad \omega_1=\hbar\,\dd\theta,
\end{equation}
where $\theta$ is the elementary loop phase. For $H_1=E_1$ we have $\dot\theta=1$. The wavefunction
\begin{equation}
    \psi_1=\psi_0\exp\!\left(\frac{\ii}{\hbar}S\right)
    =\psi_0\exp\!\left(\frac{\ii}{\hbar}E\,t\right),\quad S=\hbar.
\label{eq:psi-star}
\end{equation}
This implements a $U(1)$ phase rotation. Nevertheless, interpreting the entire universe at level $\Ch1$ as generating the electromagnetic interaction is not appropriate; the next step is required.

%========================================
\section{Transition $\C1\!\to\!\C2$: canonical pairs and metric}
%========================================

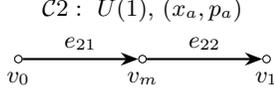
\begin{figure}[t]
\centering
\begin{tikzpicture}[scale=1.1,>=Stealth]
\node[circle,draw,inner sep=1pt,label=below:$v_0$] (A) at (0,0){};
\node[circle,draw,inner sep=1pt,label=below:$v_m$] (M) at (1.5,0){};
\node[circle,draw,inner sep=1pt,label=below:$v_1$] (B) at (3,0){};
\draw[->,thick] (A)--(M) node[midway,above] {$e_{21}$};
\draw[->,thick] (M)--(B) node[midway,above] {$e_{22}$};
\node at (1.5,0.6) {$\C2:~U(1)$, $(x_a,p_a)$};
\end{tikzpicture}
\caption{Layer $\C2$: edge splitting introduces canonical pairs and local $U(1)$ invariance.}
\label{LayerC2}
\end{figure}

\noindent According to the growth rule, the initial edge splits into two, Fig.~\ref{LayerC2}:
\begin{equation}
v_0 \xrightarrow{e_{21}} v_m \xrightarrow{e_{22}} v_1,\quad 
E_{21}+E_{22}=E_1,
\end{equation}
where $v_m$ is a new vertex. We introduce coordinates and momenta
\begin{equation}
    x_a=ct_{2a}, \quad p_a=E_{2a}/c,    
\end{equation}
with $a=1,2$ and $c$ a proportionality constant between momentum and energy. The times $t_{2a}$ are arbitrary subject to $t_{21}{+}t_{22}=2\tau_1=\tau_2$. Conservation of action allows reparametrizing times into coordinates; hence
\begin{equation}
\omega_2=\sum_{a=1}^2\dd p_a\wedge\dd x_a,
\qquad [x_a,p_b]=\ii\hbar\,\delta_{ab}.
\end{equation}

\noindent The Hamilton–Jacobi action
\begin{equation}
S[x_a]=\int(E_1\,\dd t_2-p_a\dd x_a)
\end{equation}
yields the metric
\begin{equation}
ds^2=c^2\dd t_2^2-\delta_{ab}\dd x^a\dd x^b,
\end{equation}
i.e., an effective spacetime with signature $(+--)$.

Since $S^{(2)}_{\min}=2\hbar$, common phase shifts $\theta$ act as
\begin{align}
\Phi&\mapsto\Phi+\theta,\\
E_2&\mapsto E_2-qA_0,\quad\\
p_a&\mapsto p_a-qA_a, \\
(A_0,A_a)&=\partial_{(t_2,x_a)}\theta,
\end{align}
realizing a $U(1)$ fiber over $(t_2,x_1,x_2)$. On graphs this is the Peierls phase on edges. The $(2{+}1)$ action with gravity, Abelian gauge field, and matter is
\begin{align}
S_{\text{2}}&=\frac{1}{2\kappa_2}\int d^3x\,\sqrt{|g|}\,R
-\frac{1}{4}\int d^3x\,\sqrt{|g|}\,F_{\mu\nu}F^{\mu\nu}\nonumber\\
&\quad+\int d^3x\,\sqrt{|g|}\,\mathcal{L}_{\text{matter}}(g,A,\ldots),
\end{align}
with $F=dA$ and matter covariant derivative $D_\mu=\partial_\mu+\frac{1}{4}\omega_\mu^{IJ}\gamma_{IJ}-iqA_\mu$.

In phase-locked domains of $\C2$ the bulk gravitational sector can be rewritten in Chern–Simons form. For $\Lambda=0$, the gauge field $\mathcal{A}=\omega^I J_I + e^I P_I\in\mathfrak{iso}(2,1)$ and
\begin{equation}
S_{\text{CS}}[\mathcal{A}]=\frac{k}{4\pi}\int\!\text{tr}\!\left(\mathcal{A}\wedge d\mathcal{A}
+\tfrac{2}{3}\mathcal{A}\wedge \mathcal{A}\wedge \mathcal{A}\right),
\end{equation}
impose flatness of $\mathcal{A}$, equivalent to vacuum Einstein equations in $(2{+}1)$. For $\Lambda\neq0$ the group becomes $\mathrm{SO}(2,2)$ or $\mathrm{SO}(3,1)$; matter adds Wilson lines.

%========================================
\section{Transition $\C2\!\to\!\C3$: triple closure and $SU(3)$}
%========================================

\begin{figure}[t]
\centering
\begin{tikzpicture}[scale=1.05,>=Stealth]
\node[circle,draw,inner sep=1pt,label=below:$v_0$] (A) at (0,0){};
\node[circle,draw,inner sep=1pt,label=below:$v_{m1}$] (M1) at (1,0){};
\node[circle,draw,inner sep=1pt,label=below:$v_{m2}$] (M2) at (2,0){};
\node[circle,draw,inner sep=1pt,label=below:$v_1$] (B) at (3,0){};
\draw[->,thick] (A)--(M1) node[midway,above left] {$e_{31}$};
\draw[->,thick] (M1)--(M2) node[midway,above] {$e_{32}$};
\draw[->,thick] (M2)--(B) node[midway,above right] {$e_{33}$};
\node at (1.5,1.0) {$\C3:~SU(3)$};
\end{tikzpicture}
\caption{Layer $\C3$: a triple closure yields $SU(3)$-type connections.}
\label{LayerC3}
\end{figure}
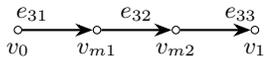

\noindent At the next step each chain closes into a triple path $(e_{31},e_{32},e_{33})$ with $E_{31}{+}E_{32}{+}E_{33}=E_1$, Fig.~\ref{LayerC3}. Define
\begin{equation}
  (A_\mu)^a{}_{b}
  \;\equiv\;
  \frac{1}{\hbar}\bigl(p_a x_a - p_b x_b\bigr)\,k_\mu^{\,a}{}_{b}.
\end{equation}
The commutator in the fundamental representation
\begin{equation}
  \bigl[ A_\mu , A_\nu \bigr]^a{}_{b}
  \;=\;
  (A_\mu)^a{}{c}\,(A\nu)^c{}_{b}
  \;-\;
  (A_\nu)^a{}{c}\,(A\mu)^c{}_{b}\,,
  \label{eq:comm-fund}
\end{equation}
is traceless and closes on $3\times3$ trace-free matrices, realizing $\mathfrak{su}(3)$. Expanding in generators $T^c$:
\begin{equation}
  A_\mu \;=\; A_\mu^c\,T^c\,,
  \qquad c=1,\dots,8\,,
  \label{eq:A-adjoint-def}
\end{equation}
\begin{equation}
  [T^c, T^d] \;=\; i f^{cde}\,T^e,
  \qquad
  \mathrm{tr}\bigl(T^c T^d\bigr)
  \;=\; \frac{1}{2}\,\delta^{cd}\,.
  \label{eq:T-comm}
\end{equation}
The maps between representations are
\begin{equation}
  (A_\mu)^a{}_{b}
  \;=\;
  A_\mu^c\,(T^c)^a{}_{b}\,,
  \qquad
  A_\mu^c \;=\; 2\,\mathrm{tr}\!\left(T^c A_\mu\right).
  \label{eq:A-fund-adjoint-map}
\end{equation}
Hence
\begin{equation}
  [A_\mu, A_\nu]
  \;=\;
  A_\mu^c A_\nu^d\,[T^c, T^d]
  \;=\;
  i f^{cde} A_\mu^c A_\nu^d\,T^e\,,
\end{equation}
and in the adjoint representation
\begin{equation}
  \bigl[ A_\mu , A_\nu \bigr]^e
  \;=\;
  i f^{e c d}\,A_\mu^c A_\nu^d\,.
  \label{eq:comm-adjoint}
\end{equation}
Introducing the standard field tensor and the action for $\C3$,
\begin{align}
F^a_{\mu\nu}&=\partial_\mu A^a_\nu-\partial_\nu A^a_\mu+g f^{abc}A^b_\mu A^c_\nu,\\
S_3&=\!\int\!\dd^3x\,\sqrt{\abs{g}}\!\left[\frac{1}{2\kappa_3}R-\frac{1}{4}F^a_{\mu\nu}F^{a\mu\nu}\right],
\end{align}
one obtains the Einstein–Yang–Mills equations. The rotational generators satisfy $[L_i,L_j]=\ii\hbar f_{ij}{}^{k}L_k$ with $J_i=L_i+S_i$.

%========================================
\section{Fourth Layer $\C4$: dimensional reduction and gauge consistency}
%========================================

\noindent The fourth layer has $\tau_4=4\tau_1=2\tau_2$ and can be viewed as $4\times\C1$ or $2\times\C2$. At energies below the layer scale these descriptions agree: $\C4$ is the minimal structure with coexisting pair locking.

Four internal times $(t_{41},t_{42},t_{43},t_{44})$ form the pairs $(1,3)$ and $(2,4)$. Strong intra-pair coupling $\kappa_2$ locks phases within each pair, while weaker inter-pair coupling $\delta$ acts across pairs. Introducing mean coordinates
\begin{equation}
X_1=\tfrac12(x_1+x_3),\qquad X_2=\tfrac12(x_2+x_4),
\end{equation}
and integrating out heavy relative modes gives an effective metric
\begin{equation}
ds_{\mathrm{eff}}^2=c^2d\tau^2-G_{ij}(X)\,dX^i dX^j,\quad i,j=1,2,
\end{equation}
so the infrared behavior of $\C4$ is that of a $(2{+}1)$ sheet.

Pair locking yields
$U(1){(13)}^{(+)}\times U(1){(24)}^{(+)}$,
and the weaker cross-pair coupling identifies the diagonal combination,
\begin{align}
&U(1){(13)}^{(+)}\times U(1){(24)}^{(+)}
\rightarrow U(1)_{\mathrm{diag}},
\\
&A_\mu^{\mathrm{diag}}=\tfrac14\sum_{a=1}^4 A_\mu^{(a)}.
\end{align}
All orthogonal combinations become massive; the surviving massless photon is identical to the $U(1)$ of $\C2$. Charge quantization and the $\mathbb Z_2$ structure are preserved. At UV scales $\C4$ behaves as a $(4{+}1)$ system; below $\kappa_2$ it reduces to two $\C2$-like doublets; below $\delta$ only the diagonal $U(1)$ remains.

%========================================
\section{Conclusion}
%========================================

\noindent The discrete-action axiom $S_1=\hbar$ and the oriented-graph growth rule suffice to build a hierarchy of layers $\C N$. Each transition $\C N\!\to\!\C{N+1}$ adds canonical pairs and strengthens symmetry: $\C2$ exhibits local $U(1)$ and a $(+--)$ metric, $\C3$ yields $SU(3)$ and the Einstein–Yang–Mills action, and $\C4$ ensures gauge consistency with dimensional reduction in the infrared. Stochastic graph growth naturally leads to decoherence and symmetry-breaking mechanisms.

%========================================
\appendix
\section*{Appendix A: Formal derivations}

\subsection{Variation of a discrete action}

\noindent Each elementary edge $e_i$ carries $\delta S_i=\hbar$. With fixed $J_z$, the variation of $S_N=N\hbar$ gives
\begin{equation}
\delta S_N=\sum_i(p_i\,\delta x_i-E_i\,\delta t_i)=0,
\end{equation}
from which the symplectic form $\omega=\sum_i\dd p_i\wedge\dd x_i$ follows.

\subsection{Emergence of canonical pairs}

\noindent At $\C2$, splitting $e_1$ doubles the action to $2\hbar$. With $E_{21}{+}E_{22}=E_1$ and $p_a=E_{2a}/c$ we obtain $[x_a,p_b]=\ii\hbar\,\delta_{ab}$—the minimal discrete analogue of phase space.

\subsection{Derivation of the $SU(3)$ structure}

\noindent At $\C3$, with phases $\phi_a$, the closure condition
\begin{equation}
\phi_{12}+\phi_{23}+\phi_{31}=0
\end{equation}
yields three independent differences forming $\mathfrak{su}(3)$:
\(
[T_a,T_b]=\ii f^{abc}T_c.
\)
Mapping $\phi_{ab}\!\to\!A_\mu^{(ab)}T_{ab}$ leads to the curvature
\(
F=\dd A + g A\wedge A.
\)

%========================================

\end{document}